# Diffusive Topological Transport in Spatiotemporal Thermal Lattices

Guoqiang Xu[1], Yihao Yang[2], Xue Zhou[3], Hongsheng Chen[2,4,5], Andrea Alù[6,7], Cheng-Wei Qiu[1*]

[1]Department of Electrical and Computer Engineering, National University of Singapore, Kent Ridge 117583, Republic of Singapore
[2]Interdisciplinary Center for Quantum Information, State Key Laboratory of Modern Optical Instrumentation, ZJU-Hangzhou Global Scientific and Technological Innovation Center, Zhejiang University, Hangzhou, China
[3]School of Computer Science and Information Engineering, Chongqing Technology and Business University, Chongqing, 400067, China
[4]ZJU-Hangzhou Global Science and Technology Innovation Center, Key Lab. of Advanced Micro/Nano Electronic Devices & Smart Systems of Zhejiang, Zhejiang University, Hangzhou 310027, China
[5]International Joint Innovation Center, ZJU-UIUC Institute, The Electromagnetics Academy at Zhejiang University, Zhejiang University, Haining 314400, China
[6]Photonics Initiative, Advanced Science Research Center, City University of New York, New York, New York 10031, USA
[7]Physics Program, Graduate Center of the City University of New York, New York, New York 10016, USA

*Corresponding author. Email: chengwei.qiu@nus.edu.sg

**Abstract:** Topological insulating phases are usually found in periodic lattices stemming from collective resonant effects, and it may thus be expected that similar features may be prohibited in thermal diffusion, given its purely dissipative and largely incoherent nature. We report the diffusion-based topological states supported by spatiotemporally-modulated advections stacked over a fluidic surface, thereby imitating a periodic propagating potential in effective thermal lattices. We observe edge and bulk states within purely nontrivial and trivial lattices, respectively. At interfaces between these two types of lattices, the diffusive system exhibits interface states, manifesting inhomogeneous thermal properties on the fluidic surface. Our findings establish a framework for topological diffusion and thermal edge/bulk states, and it may empower a distinct mechanism for flexible manipulation of robust heat and mass transfer.

Topological insulating phases have been unveiling a variety of new wave phenomena in metamaterials. A particularly interesting class of topological phenomena arises in open systems (*1, 2*), enabling novel phenomena across a variety of fields (*3-7*), such as anomalous edge states (*8, 9*), mode switching (*10*), unidirectional wave propagation (*11, 12*), single-mode laser (*13*), pump-dependent lasing (*14-16*), etc. Two main recipes have been proposed to realize these classes of non-Hermitian topological responses. The first option is to drive the topological responses by magneto-optic effects (*17*) or tailored lattice coupling (*18*) in cooperation with the introduced non-Hermitian elements; the second is to induce nontrivial topological phases by pumping non-Hermitian elements like gain and loss (*19-25*) into the otherwise Hermitian systems.

Recently the Hamiltonian associated with classical thermal diffusion was discovered to follow a skew-Hermitian relation (*26*), which may offer a hint to realize a topological insulating phase for thermal diffusion. On the other hand, the absence of a periodic potential and dynamic coherence for thermal transport fundamentally hinders this pathway. To synthesize a topological insulating phase in purely dissipative diffusion mechanism, it is of utmost importance to imitate the periodic potential and coherent interference in thermal transport, such that Bloch theory could be revived and diffusive topological transport could be thereby expected as those photonic and acoustic counterparts (*17-25*).

Here, we resort to judiciously time-modulated Hermiticity features in both space and time to introduce an advective paradigm for the very first demonstration of thermal topological transport in periodically stacked fluid surfaces. We incorporate spatiotemporally modulated advections to enable an additional real dimension to thermal diffusion, while alternating advection arrangements offer the imitated coherences and periodic potentials. Depending on the advective configuration, the effective bandgap can be either topologically nontrivial or trivial. We experimentally discover thermal topological modes with edge and interface states as well as conventional bulk states, respectively manifested as stationary and deviated temperature distributions. The findings provide a feasible way of creating the effective periodic lattice on an advective fluidic surface, unlocking more opportunities and applications for topological thermal diffusion. Our proposed recipe for diffusive topological modes can further enlighten the manipulation of general diffusive fields (*27-31*).

We consider a fluidic surface possessing periodic advections, as shown in Fig. 1**A**, to demonstrate non-Hermitian topological modes for thermal diffusion. Owing to the dissipative nature of thermal diffusion, non-Hermiticity inherently exists in the proposed system. In order to create an effective coherent resonance for thermal fields over the fluidic surface, we apply spatiotemporal modulation onto corresponding advective regions (*32*) marked by specific colors (Fig. 1**A**). These advective blocks could be regarded as four modulated units, forming one effective lattice. Further periodically configuring multiple effective lattices, an effective 1D chain can be induced on the fluidic surface. Here, the heat transfer process in one super cell of the effective lattice can be written as

$$\rho c \nabla_t T_n = \kappa \nabla_x^2 T_n \pm \rho c v_n \nabla_x T_n + \frac{h_n}{a_n}(T_{n+1} - T_n). \tag{1}$$

In Eq. (1), $\rho$, $c$ and $\kappa$ respectively denote the density, specific heat, and thermal conductivity of the fluid domain. $T_n$ is the temperature of a specific unit of one lattice shown in Fig. 1**A**, while the subscript $n$ indicates the unit index counting from 1. $a_n$ and $h_n$ are the width and convective heat transfer coefficient of each unit. For simplification, we adopt the same width for each unit, thus leading to an effective lattice constant of 4$a$. We further assign a series of spatiotemporal advections ($v_i$, -$v_{ii}$, -$v_i$, $v_{ii}$) to each unit advection ($v_n$, $v_{n+1}$, $v_{n+2}$, $v_{n+3}$) of one lattice to create effective "oscillation" in such diffusive systems, i.e., $\pm \rho c v_n \nabla_x T_n$ in Eq. 1, where the sign indicates the advective direction in each units and $v_i$, $v_{ii} \geq 0$. Owing to the spatiotemporal properties, these advective velocities imitate the periodic potential fields yielding $v_n = v_{n+4}$ between two adjacent lattices (Eq. S2). Upon the interaction of imposed advections and conductions, their thermal exchange $\pm \frac{h_n}{a_n}(T_{n+1} - T_n)$ effectively provides the coupling within the neighboring units. The phase diagram for the effective band structure under specific advective configurations is presented in Fig. 1**B**, indicating four insulating phases I ~ IV (Supplementary Note 2). The thicknesses of the advective components and the target fluidic surface are respectively $b$ and $d$, as presented in Fig. 1**C**. Under the hypothesis of small fluid thermal conductivity $\kappa$, the first-order wave-like solution $T_n = Ae^{i(k_n x - \omega_n t)}$ hints the possibility of switching Eq. (1) to a similar form of the Schrödinger equation, where $k_n = \frac{2\pi}{L_n} = R_n^{-1}$ and $\omega_n = -i\frac{2\kappa \cdot k_n}{\rho c} - k_n v_n$ denote the effective wave numbers and the angular frequencies of units for each lattice. Owing to the same radii of each unit, the effective wavenumbers are same for the entire system, i.e., $k_n = k = R^{-1}$. The changing angular velocity corresponds to a real angular frequency ($\omega_n$) in the thermal system. Based on Bloch theorem and the imposed periodic potential (velocity) field, the effective Hamiltonian can be expressed as

$$H = i \cdot \begin{bmatrix} i \cdot kv_i & \dfrac{h}{\rho cb} & 0 & \dfrac{h}{\rho cb}e^{ik_t \cdot 4a} \\ \dfrac{h}{\rho cb} & -i \cdot kv_{ii} & \dfrac{h}{\rho cb} & 0 \\ 0 & \dfrac{h}{\rho cb} & -i \cdot kv_i & \dfrac{h}{\rho cb} \\ \dfrac{h}{\rho cb}e^{-ik_t \cdot 4a} & 0 & \dfrac{h}{\rho cb} & i \cdot kv_{ii} \end{bmatrix} - i \cdot \left( \dfrac{\kappa}{\rho c} \cdot k^2 + \dfrac{h}{\rho cb} \right) \cdot I_{4\times 4}. \quad (2)$$

where $k_t$ denotes the effective Bloch wave number (Supplementary Note 1). Here, $D = \dfrac{\kappa}{\rho c}$ denotes the diffusivity, while we further assume $m = \dfrac{h}{\rho cb} = \dfrac{\kappa}{\rho cbd}$ for simplification under the hypothesis of small thickness of the fluidic surface $d$ (Fig. 1**C**). The mathematical presence of '$i$' in the entire Hamiltonian in Eq. (2) implies that the proposed diffusive system is orthogonal to the Hamiltonian of photonic systems. Thus, the imposed advections act as a Hermitian modulation in this diffusion scenario, and it is essentially equivalent to the non-Hermitian role played by gain and loss in the photonic counterpart (*19, 22-25*). Thus, we can define eigenvalues for the periodic lattice

$$E_{\pm} = -i\left(D \cdot k^2 + m\right) \pm \dfrac{\sqrt{2}}{2}\sqrt{p \pm \sqrt{p^2 - q^2 - l^2}}.$$

$$p = (kv_i)^2 + (kv_{ii})^2 - 4m^2, q = 2(kv_i)(kv_{ii}), s = m^2\left(1 + e^{ik_t a}\right), l = 4m^2 \sin\left(\dfrac{k_t a}{2}\right). \quad (3)$$

This system can support topological modes, since Eq. (2) obeys pseudo-anti-Hermiticity, i.e., $H = -\eta H^{\dagger} \eta$, where $\eta = diag(1,-1,1,-1)$. Considering one four-unit lattice, two types of advective couplings can be expected in Eq. (3) with imposed velocities ($v_i$, $-v_{ii}$, $-v_i$, $v_{ii}$), thus leading to effective couplings $\sqrt{\left(\dfrac{h}{\rho cb}\right)^2 - (kv_n \pm kv_{n+1})^2}$ between any two adjacent units. These advective couplings within the four-unit lattice support an effective bandgap $\Delta = \sqrt{p - \sqrt{p^2 - q^2}}$ and the four insulating phases shown in Fig. 1**B**. Here, we focus on the response in the gapped phase III with $p - q > 0$ and $p < -q$. In this phase, the integer winding number is nonzero/zero when the value $v_i v_{ii}$ is larger/smaller than 0. The other insulating phases and corresponding properties are discussed in Supplementary Note 2.

**Topologically nontrivial and trivial response of the finite lattice**
Nonzero and zero winding numbers respectively correspond to the presence or absence of edge states at the boundaries. In order to verify their presence, we consider 40 units forming 10 effective lattices in the schematic model shown in Fig. 1**C** to create a finite fluidic system. The inherent conduction between each pair of adjacent lattices acts as the hopping term to generate the underlying edge and bulk states. When we consider $v_i > 0$ and $v_{ii} > 0$ for each lattice in the selected phase, the sorted angular velocities (eigenfrequencies) indicate the emergence of a pair of topological edge states within the bandgap, as illustrated in the upper insert of Fig. 1**D**. For $v_i > 0$ and $v_{ii} < 0$ in the same phase, only the bulk state emerges in the lower insert of Fig. 1**D**. The thermal distribution of the edge and bulk states in the selected phase are illustrated in Fig. 2. The captured temperature profiles at specific moments of the nontrivial lattice scheme are presented in Fig. 2**A**. The thermal behaviors are robust, with stationary temperature profiles during the entire thermal process, while some deviations relative to initial locations can be observed at the two boundaries of the system. Further tracing the experimental locations of the maximum temperature ($T_{max}$) of each unit on the fluidic surface, we find that they correspond to the first and last lattices at the two system boundaries respectively, and slightly move to the two sides along $y$-direction revealing the nonzero wining number ($\pm 1$) (Supplementary Note 2). The $T_{max}$ locations of the other units almost remain unchanged (Fig. 2**C**). Then, we study the state intensities of each unit, which could be described as the effective thermal resistance in analog to the topological electric circuits (20) respectively along $z$-direction between two adjacent units and in the $x$-$y$ plane within one unit, i.e., $R_{eff,z} = \frac{T_{n+1,z}-T_{n,z}}{P} = \frac{\Sigma(\psi_{n+1}-\psi_n)}{E_\pm}$ and $R_{eff,xy} = \frac{\Delta T_{n+1,xy}}{P}$, where $T_{n+1,z}$ and $T_{n,z}$ are the temperature components along $z$-direction of adjacent lattices, and $\Delta T_{n+1,xy}$ is the temperature difference component between the highest and lowest temperatures in the $x$-$y$ plane of a specific unit. Since the eigenvalue reveals the energy per unit of time corresponding to the effective heat flux along the direction of lattice arrangements, each pole of $R_{eff,z}$ and $R_{eff,xy}$ can be used to indicate the corresponding modes on the fluidic surface. Thus, the larger $R_{eff,z}$ and smaller $R_{eff,xy}$ at the system boundaries compared to the other regions shown in Fig. 2**C** and **E** hint to the existence of edge states (the deviations observed in Fig. 2**A**), which respectively suppress and contribute to the heat flux propagation towards specific directions. It is worth noting that the stationary thermal distribution can be observed over the entire system, owing to the thermal equilibrium under such advective configurations. The smaller $R_{eff,z}$/larger $R_{eff,xy}$ (not at the boundaries) lead to expedite/tough heat transfer towards these directions.

In contrast, the thermal features of the scheme with trivial lattices (the lower insert of Fig. 1**D**) are presented in Figs. 2**B**, **D**, and **F**. The temperature profiles and $T_{max}$ locations shown in Fig. 2**B** indicate that the thermal distributions exhibit distinctive differences between initial and final moments. The characteristic thermal distributions showcase significant deviations towards $\pm y$ directions in the units near both system boundaries and reveal relatively stable distributions in the intermediate units of the system over time. Such behaviors are robustly maintained till steady state. Further evaluating the state intensities (effective thermal resistances along $z$-direction and in the $x$-$y$ plane) of each unit for the trivial lattices, there are no significant differences between the effective thermal resistances of each unit in these directions, revealing the bulk state on the fluidic surface with zero winding number (Supplementary Note 2). Except the different characteristics of the temperature distributions, the thermal process of the scheme with trivial lattices is enhanced compared to the nontrivial one, since the temperature amplitudes shown in Fig. 2**B** indicate a faster relaxation process than the one of Fig. 2**A**.

**Topological manipulation of thermal profile**
The topological edge and bulk states suggest the possibility of flexibly manipulating the thermal properties of a fluidic surface. Here, we further explore two cases with tunable butting of topologically nontrivial and trivial lattices, respectively possessing the nonzero ($\pm 1$) and zero winding numbers, to demonstrate controllable edge, interface and bulk states, while the advections are also modulated in the same gapped phase (Supplementary Note 2). For Case I, we respectively configure five nontrivial lattices (20 units) and five trivial lattices (20 units) to the fluidic surface as shown in Fig. 3**A**. The characteristic thermal distributions and experimental temperature profiles are respectively illustrated in Figs. 3**B** ~ **D**. The $T_{max}$ locations at specific moments (Fig. 3**B**) indicate that the characteristic distributions of the nontrivial lattices could maintain stationary locations in most of the units, while some deviations to the initial states occur in the units approaching the boundary of the nontrivial and trivial lattices. For the trivial lattices, the characteristic distributions showcase large and approximate deviations to the initial states in most of the intermediary units, while the $T_{max}$ locations of the last few units approaching the system boundary further indicate near-$\pi$ deviations. Such distributions in the nontrivial and trivial lattices reveal one edge state and one interface state respectively at the system boundary in the nontrivial lattice and the butting of the two lattices, implying larger effective thermal resistance as presented in Fig. 3**C**. The captured temperature profiles shown in Fig. 3**D** further validate the above characteristic $T_{max}$ locations and demonstrate distinctive thermal distributions of the two lattices. Besides, the faster relaxation can be also observed in the trivial lattices with the significant homogenized temperature amplitudes at different moments.

For Case II, we adjust the configuration strategy by respectively inserting one nontrivial lattice (4 units) at the two system boundaries, while eight trivial lattices (32 units) are further filled in the regions between the two nontrivial lattices as illustrated in Fig. 4**A**. The characteristic distributions of each unit (Fig. 4**B**) indicate that the $T_{max}$ locations of the two nontrivial lattices at the system boundaries almost remain unchanged during the measured thermal processes, while the ones of trivial lattices exhibit significant deviations to the initial states. The experimental temperature profiles at specific moments presented in Fig. 4**B** further manifest the thermal behavior in these lattices, revealing almost stationary profiles in the two nontrivial lattices and largely deviated distributions in the trivial lattices. Besides, faster relaxations can be also indicated in the trivial lattices with quicker temperature homogenizations than the nontrivial ones. Further evaluating the state intensities with the effective thermal resistance (Fig. 4**C** and **D**), two edge states at both system boundaries and two interface states at the butting of nontrivial and trivial lattices are observed with larger (smaller) effective thermal resistances along $z$-direction (in the $x$-$y$ plane) in comparison to the units of the trivial lattices. The above two cases reveal the possibility of flexibly manipulating thermal properties on a fluidic surface with the nontrivial and trivial lattices, and more modulations can be anticipated with different lattice configurations and advections.

This work introduces an advective route to reveal for the first-time thermal topological modes by stacking periodic advections on a fluid surface. These spatiotemporal-modulated advections introduce manipulation in momentum space for a purely dissipative phenomenon like thermal diffusion, enabling effectively nontrivial and trivial lattices. Thermal topological edge, interface, and bulk states were observed upon modulating the advections within these effective lattices in a gapped phase. Our findings point to a direction beyond conventional topological physics based on oscillatory fields, and offer a unique platform to study unexpected non-Hermitian topological modes in pure diffusive systems. Besides, this work also suggests a distinctive mechanism of arbitrary diffusive manipulations with the combinations of nontrivial and trivial lattices.

**Acknowledgments:**

**Funding:**

Ministry of Education, Republic of Singapore, Grant No.: R-263-000-E19-114 (CWQ).

**Author contributions:**

Conceptualization: GX, CWQ

Methodology: GX, XZ, CWQ

Investigation: GX, YY, HC, AA, CWQ

Visualization: GX, YY, XZ, HC, AA, CWQ

Funding acquisition: CWQ

Project administration: GX, CWQ

Supervision: CWQ

Writing – original draft: GX, CWQ

Writing – review & editing: GX, YY, XZ, HC, AA, CWQ

**Competing interests:** Authors declare that they have no competing interests

**Data and materials availability:** All data are available in the main text or the supplementary materials.

**Supplementary Materials**

Supplementary Text

Materials and Methods

Figs. S1 to S5

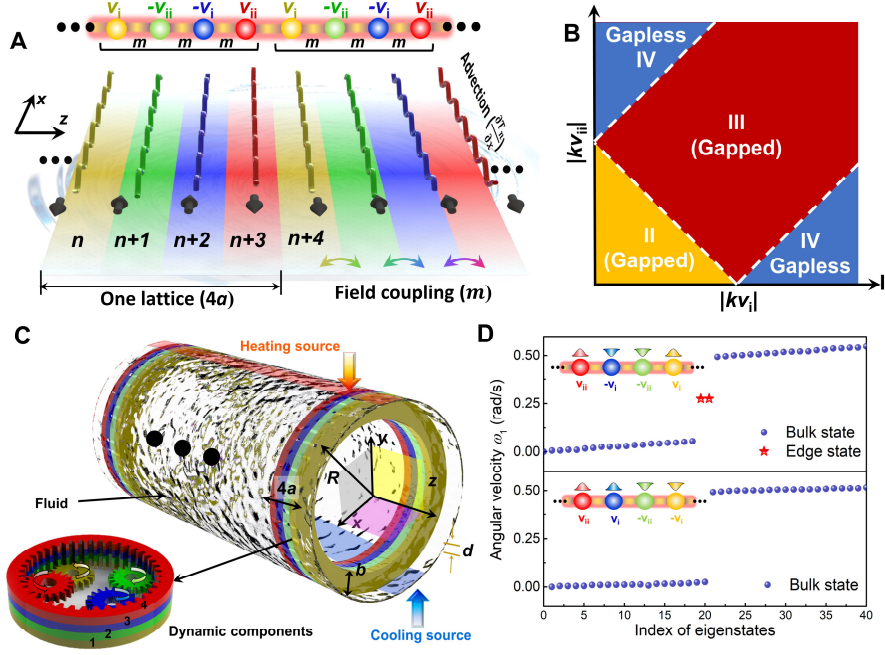

**Fig. 1. Topological transport in thermal diffusion and experimental setups. A.** Schematic of realizing topology in thermal diffusion by introducing periodic advections onto a fluidic surface (colored curves). These periodic advections are independently imposed, thus contributing to the effective "lattice" with four units marked by n~ n+3 on the fluidic surface. The lattice constant is $4a$, while the field coupling between two adjacent units is described as $m = \frac{h}{\rho c b} = \frac{\kappa}{\rho c b d}$. **B** indicates the phase diagram for the effective band structure. Among these phases, Phase I exists only when $v_i = 0$ or $v_{ii} = 0$; Phases II and III are gapped, while Phase IV is gapless. **C** illustrates the experimental setup for observing the thermal topological transport on a fluidic surface. The light red and light blue strips on the top and bottom respectively denote the heating and cooling sources for providing the initial temperature field. The thermal process is modulated by ten effective lattices as shown in the lower insert. Each effective lattice consists of four dynamic units for providing specific advections. **D** presents the angular velocity (eigenfrequencies) sorted in a gapped phase as a function of $kv_i = m$, $kv_{ii} = 2m$. The upper and lower inserts respectively indicate the nontrivial and trivial lattices.

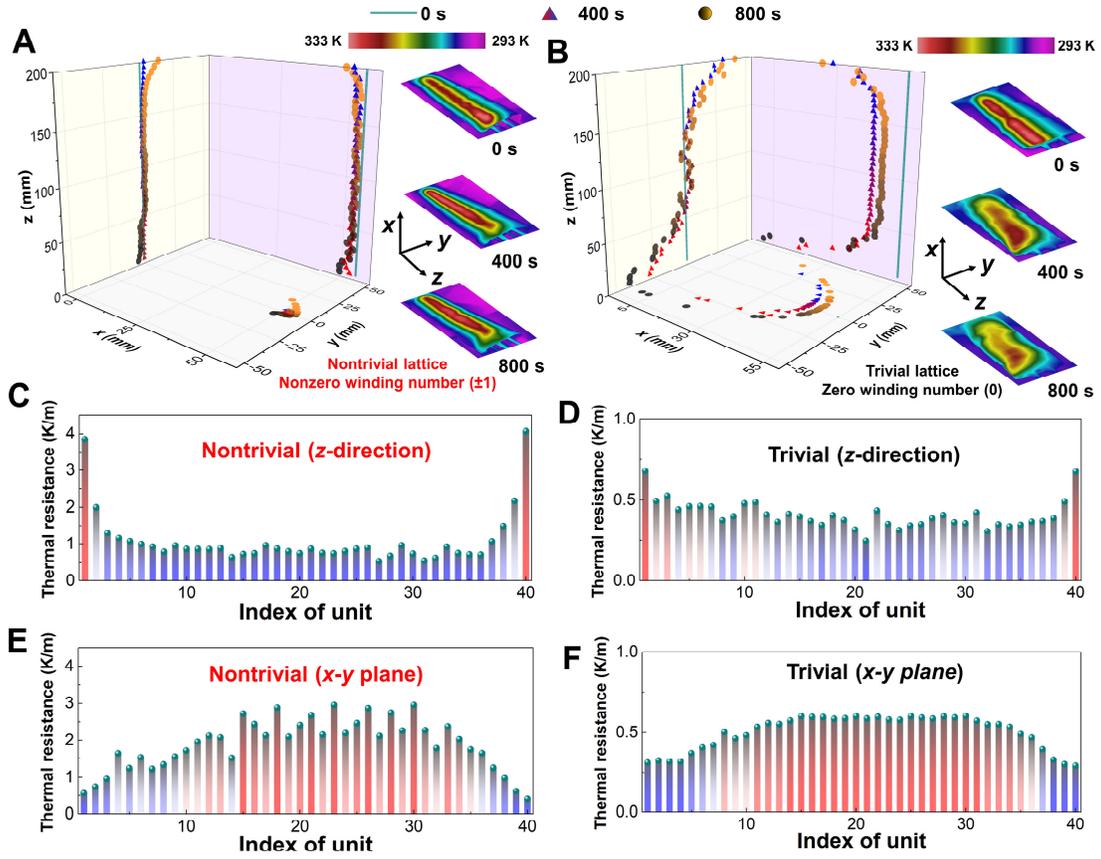

**Fig. 2. Measured results for topologically nontrivial and trivial responses in the gapped phase.**
**A** and **B** respectively present the experimental $T_{max}$ locations (the left panel) and the thermal images (the right panel) of the nontrivial and trivial lattices measured at specific moments. The temperature profile on the fluidic surface keeps well confined along *y*-direction at nontrivial lattice situation (**A**), but it expands to both sides along *y*-direction under the condition of trivial lattice (**B**). **C** and **D** provide the sums of the amplitudes of the effective thermal resistance along *z*-direction for the system respectively with the above two lattices. **E** and **F** illustrate the effective thermal resistances in the *x-y* plane respectively for the two lattices.

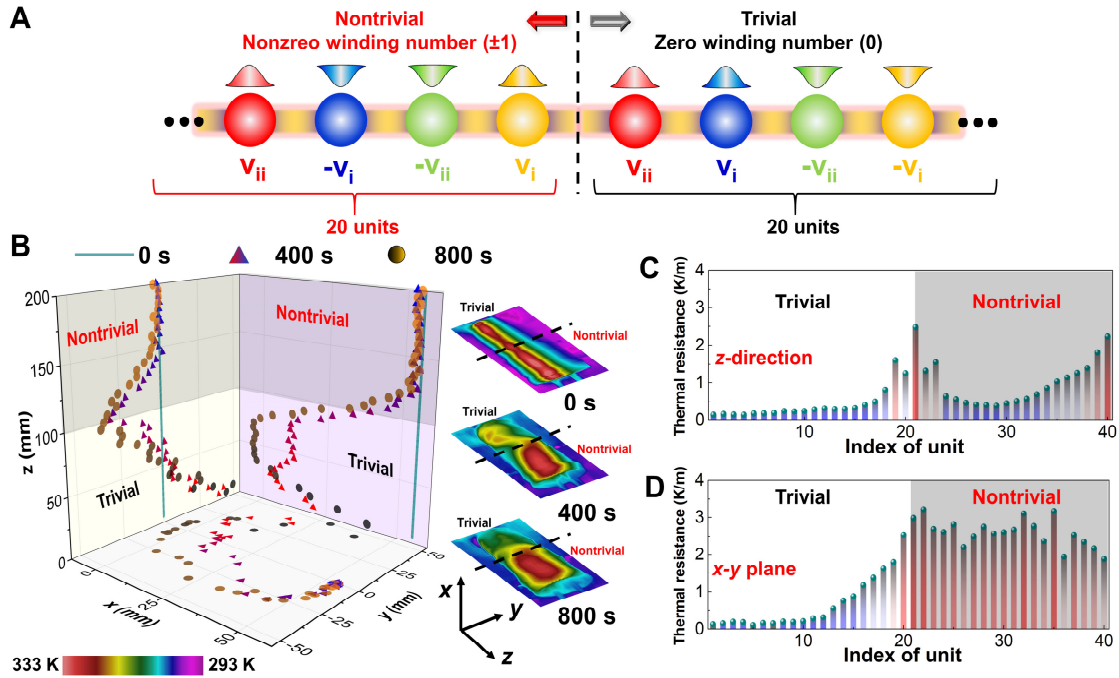

**Fig. 3. Measured temperature profiles and effective thermal resistances of Case I.** **A** denotes the advective configurations of Case I, consisting of five nontrivial lattices (20 units) and five trivial lattices (20 units). All the advections are selected in Phase III (Supplementary Note 2). **B** presents the $T_{max}$ locations (the left panel) and the captured temperature profiles (the right panel). The characteristic distributions of the nontrivial lattices almost keep stationery, while the ones of the trivial lattices move as the times go by and exhibit the largely deviated profiles. **C** and **D** indicate the effective thermal resistances Case I respectively along $z$-direction and in the $x$-$y$ plane. An interface state can be revealed at the butting between the nontrivial and trivial lattices, while an edge state is also significant at the boundary between the nontrivial lattice and open ambient. The grey shadows in **B** ~ **D** imply the regions of nontrivial lattices.

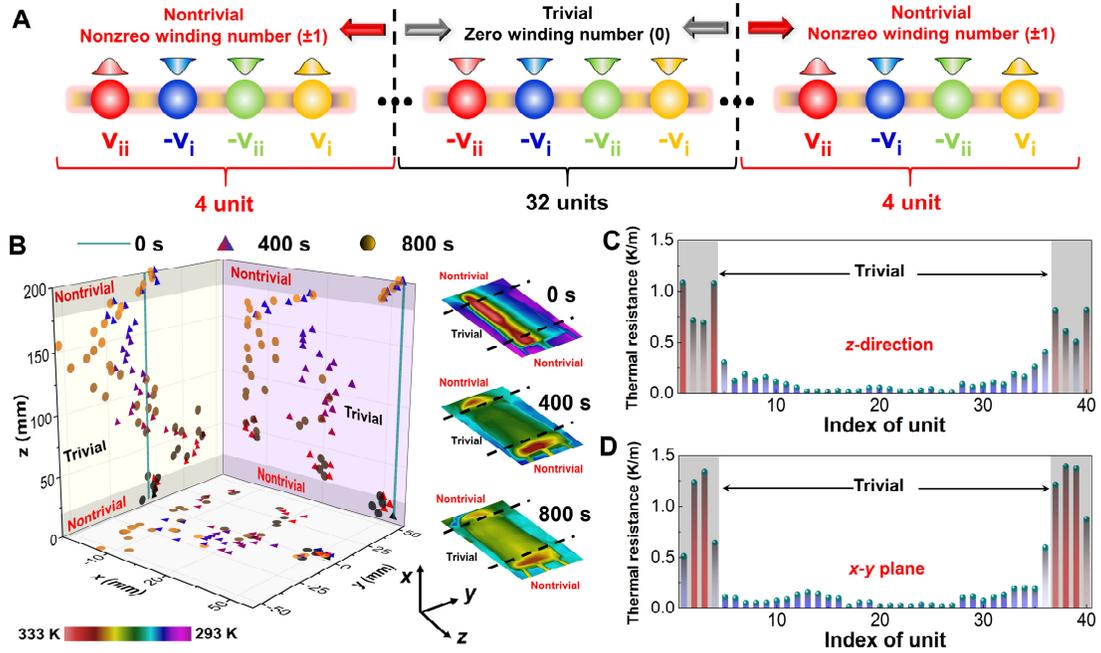

**Fig. 4. Measured temperature profiles and effective thermal resistances of Case II. A** presents the advective configurations of Case II, consisting of two nontrivial lattices respectively on both sides of the entire model, and eight trivial lattices (32 units) between the nontrivial ones. All the advections are selected in Phase III. **B** presents the $T_{max}$ locations (the left panel) and the captured temperature profiles of Case II (the right panel). Stationary thermal distributions are significant in the two nontrivial lattices at the system boundaries, while largely deviated profiles are observed in the rest trivial lattices. **C** and **D** present the effective thermal resistances of Case II respectively along *z*-direction and in the *x-y* plane. Two edge states and two interface states are observed respectively at both boundaries of the entire system and the two joints between the nontrivial and trivial lattices. The grey shadows in **B** ~ **D** indicate the regions of nontrivial lattices.